\documentclass[preprint]{ptephy_v1}

\preprintnumber{XXXX-XXXX} 
\usepackage{hyperref}
\usepackage{graphicx}
\usepackage{dcolumn}
\usepackage{bm}
\usepackage[utf8]{inputenc}
\usepackage{array}
\usepackage{makecell}
\usepackage{nicematrix}
\usepackage{multirow}
\usepackage[caption=false]{subfig}
\usepackage{amsmath}
\usepackage{amssymb}
\usepackage{float}

\begin{document}

\title{Free Particle to Complex KdV breathers through Isospectral Deformation}


\author{Kumar Abhinav\thanks{Corresponding Author}}
\affil{Centre for Theoretical Physics and Natural Philosophy,\\Mahidol University, Nakhonsawan Campus,Phayuha Khiri, Nakhonsawan 60130, Thailand
\email{kumar.abh@mahidol.ac.th}}

\author{Aradhya Shukla}
\affil{Department of Physics, Institute of Applied Sciences and Humanities,GLA University, Mathura, Uttar Pradesh 281406, India \email{ashukla038@gmail.com}}

\author{Prasanta K. Panigrahi}
\affil{Indian Institute of Science Education and Research Kolkata,\\Mohanpur, West Bengal 741246, India \email{pprasanta@iiserkol.ac.in}}


\date{\today}

\begin{abstract}%
The free particle in quantum mechanics in real space is endowed with supersymmetry, which enables a natural extension to complex spectra with a built-in parity (${\cal P}$) and time reversal (${\cal T}$) symmetry. It also explains the origin of unbroken and broken phases of the $\mathcal {PT}$-symmetry and their relationship with the real and complex eigenvalues respectively, the latter further displaying zero-width resonances. This is possible as the extension of the eigenvalue problem to the complex plane enables the incorporation of bound and decaying states in the enlarged Hilbert space. The inherent freedom of modification of the potential without changing the spectra in supersymmetry naturally explains the connection of complex breather solutions of KdV with $\mathcal {PT}$-symmetry and the free particle on the complex plane. Further, non-trivial zero-width resonances in the broken $\mathcal {PT}$ phase mandate a generalization that is directly connected to the $sl(2,\mathbb{R})$ potential algebra.
\end{abstract}

\subjectindex{05.45.-a, 03.65.-w, 11.30.Er}
\maketitle

\section{Introduction}

Supersymmetric quantum mechanics (SUSY QM) \cite{witten, khare} connects two independent Hamiltonians having the same energy spectra except for the ground state \cite{SQM}. Consequently, even a trivially solvable potential may be connected to a non-trivial potential and vice-versa. The latter fact provides a convenient algebraic way to completely solve a complicated Schr\"odinger equation. For example, the free particle has a supersymmetric partner potential that can either be periodic Scarf or finite depth P\"oschl-Teller potential \cite{grant}. Further, for a given potential, the supersymmetry algebra allows for isospectral deformation \cite{khare} that generates a class of partner potentials with identical spectra \cite{khare1, jen}. In addition, starting from the P\"oschl-Teller potential \cite{poschl}, isospectral deformation \cite{Khare2} can lead to soliton \cite{cnk} and multi-soliton solutions of the non-linear Korteweg–de Vries (KdV) equation \cite{kdv1,kdv2,kdv3} which are applicable to various physical phenomena. The multi-soliton solutions have multiple independent `time' parameters, whose number matches with the number of the bound states of the potential \cite{ SQM, sukh, pkp01,pkp02}. This owes to the fact that the Schr\"odinger eigenvalue equation corresponds to one of the Lax pairs \cite{lax1,lax2} whose compatibility leads to the integrable KdV equation \cite{miura1}. Inverse scattering transformation provides further justification for this connection \cite{cv1,cv2,cv3}. The reflectionless property of the P\"oschl-Teller potential, a consequence of its isospectrality to free particle, naturally explains the observed behavior of the solitons passing through each other without scattering \cite{lek}. Apart from these localized ones, the KdV equation is known to possess singular solitons \cite{hugo}, complex solitons \cite{buti, MSP, correa}, periodic solutions \cite{nov11,nov12} and breather solutions \cite{zaw, chaos}. Interestingly, the complex solitons of the KdV equation have been found to be invariant under parity ($\mathcal{P}$) and time-reversal ($\mathcal{T}$) symmetries \cite{MSP, correa}. As for the breather solutions, there are two types: Akhmediev breathers \cite{akh11,akh12,akh13} which are periodic in space and localized in time and Ma breathers \cite{ma} which have the opposite behavior. Under certain limited conditions, breathers tend to generate rogue waves \cite{akh31,akh32}. 

The $\mathcal {PT}$-symmetric systems require the real and imaginary parts of potential to respectively be even and odd under parity \cite{BB98a,BB98b,BB98c,BB98d,BB98e}. It is worth noting that the $\mathcal{PT}$-symmetric complex potential accommodates real energy spectra until a certain parametric threshold above which the spectrum constitutes complex-conjugate pairs \cite{lev, ZAhmed, bender}, representing the spontaneous breaking of $\mathcal {PT}$-symmetry \cite{GSD1,GSD2,GSD3,GSD4}. Recently, for the complex generalization of P\"oschl-Teller potential, a parametric interconnection between the broken and unbroken phases of $\mathcal {PT}$ and SUSY has been established \cite{pkp1}. Therein zero-width resonances \cite{ZA11, most03a,most03b}, having  real energy, are found in the broken $\mathcal {PT}$ phase \cite{yu}. A conserved non-local correlation has been shown to exist in $\mathcal{PT}$-symmetric systems with counter-intuitive implications on scattering \cite{KAP} that were experimentally observed \cite{WanW}, with a suitable `norm' distinct from the von Neumann-Dirac case of standard quantum mechanics \cite{bender, KAP, sp11,sp12}. In the unbroken $\mathcal{PT}$ phase, the complex potential and its corresponding superpotential satisfy the KdV and modified KdV (mKdV) equations respectively \cite{MSP, pkp1} under certain parametric conditions. Both periodic and localized solitons of KdV have been related to the complex P\"oschl-Teller potential \cite{buti, MSP}. It is, therefore of deep interest, to investigate if the complex $\mathcal{PT}$-symmetric breathers \cite{correa} can also be related to free particles through isospectral deformation. Further, the free particle connection of the broken $\mathcal{PT}$ phase of P\"oschl-Teller potential, where the energy is complex, is of significant interest as it will throw light on its physical origin.

A systematic study of $\mathcal {PT}$-symmetric P\"oschl-Teller potential in non-linear systems reveals an interplay between SUSY and non-linearity for the existence of stable solitons \cite{kev1}.
Remarkably, the stable solitons have been observed in $\mathcal {PT}$-symmetric optical lattice \cite{wim}, where the intrinsic non-linearity of the system plays an important role in their stability \cite{alex}.
In addition, the non-linear systems possessing  $\mathcal{PT}$-symmetry have been the subject of many recent investigations \cite{fring11,fring12,fring13,fring14,fring15,fring16}. The effect of non-linearity on $\mathcal{PT}$-symmetric potentials has been shown to lead to self-trapped modes \cite{mu1,mu2}. Moreover, a system having defocusing non-linearity and odd gain-loss distribution has been found to have real energy spectra \cite{malo01} owing to its $\mathcal{PT}$-symmetry that becomes unbreakable for arbitrarily large strength of gain-loss term \cite{malo02}.  Further, the notion of supersymmetry in  $\mathcal {PT}$-symmetric non-linear dual-core system has been found to yield solitonic solutions \cite{malo03}, owing their stability to the periodic change in the sign of gain-loss and dual-core coupling \cite{malo04}. Interestingly, for this system, the initial discrepancy in amplitude and phase between the two modes transform the soliton into a breather solution \cite{malo05}. 

It is well-known that the KdV and mKdV equations are related through Miura transformation:  $u = v^2 \pm v_x$,  \cite{miura2, kr02}, where $u(x,t)$ and $v(x,t)$ are solutions of KdV and mKdV equations, respectively. In addition, the existence of two distinct mKdV equations, whose solutions $v(x, t)$ are connected by $v \rightarrow iv$, generate another class of solutions having the functional form: $u = -v^2 \pm iv_x$, which also satisfy the KdV equation \cite{buti, miura2}.  This is significant as it suggests complexification of the space for a better understanding of the complex potential and their connection with broken and unbroken phases of $\mathcal{PT}$.

For the demonstrative purpose, herein we employ isospectral deformation to a free particle in an infinite potential well having a general boundary condition. Although simple and well-documented \cite{Khare2}, the present treatment elucidates the crucial impact of the coordinate boundaries on and subsequent scope for generalization of the ground state, which remains implicit in the usual Schr\"odinger differential equation approach. Consequently, the most general quantum system is needed to be defined on the complex plane with the `physical' unidirectional motion identified parallel to the real axis. The `general' isospectral deformation subsequently obtained effects a {\it complex} shift, mapping to a complex P\"oschl-Teller system having $\mathcal{PT}$-symmetry, wherein parity is defined in real space and the imaginary coordinate being a continuous parameter.
 We adhere to this definition of parity in order to remain close to physical realization. It follows that the width of the well, which has to be real for the realness of the physical spectrum, has to be an integral multiple of a constant. The emergent $\mathcal{PT}$-symmetric system, having well-separated real and imaginary parts, describes the behavior of the breather solution in the corresponding KdV system. Under a different parametric identification, this potential also leads to a complex soliton solution of the KdV equation. A different class of parameterization further yields both breather and complex soliton solutions that also satisfy the Miura transformation equations, thereby identifying the corresponding superpotential as solutions to the mKdV equation. Interestingly, with a further extension of the ground state that conforms to an underlying $sl(2,\mathbb{R})$ potential algebra, a more general  $\mathcal{PT}$-symmetric system is obtained possessing a spontaneously $\mathcal{PT}$-broken sector with complexc-onjugate spectra that includes zero-width resonance states. It is to be noted that, the complex potential in the broken-$\mathcal{PT}$ phase does not satisfy the KdV equation and its connection with the free particle is not possible.   

The paper has been organized as follows. Section \ref{sec:level2} contains a brief summary of SUSY QM and isospectral deformation through the solutions of  Bernoulli's equation. Section \ref{sec:level3} deals with the family of equivalent superpotentials corresponding to 1-D particles in an infinite box and resulting isospectral family of potentials, necessarily requiring complexification of space and discretization of the width of the well. Remarkably, the emergent  potentials remain invariant under $\mathcal{PT}$-symmetry and are exact breather solutions of the KdV equation. Section \ref{sec:level5} analyzes the $\mathcal{PT}$-symmetric potentials, isospectrally constructed from the generalized particle in a box, and also the general parametric conditions leading to spontaneous breaking of $\mathcal{PT}$-symmetry. Section \ref{sec:level6} concludes the paper along with the prospects for future works.

\section{\label{sec:level2} Intertwined Hamiltonians and isospectral deformation}
Supersymmetric quantum mechanics relates the two Hamiltonians\footnote{We take $\hbar = 2m = 1$ in the whole text.},
\begin{eqnarray}
H_{\pm}(x) = -\frac{d^2}{dx^2} + V_{\pm}(x), 
\end{eqnarray}
 with supersymmetric partner potentials,
 \begin{eqnarray}
 V_{\pm}(x) = W^2(x) \pm W'(x).\label{E2}
 \end{eqnarray}
defined through a common {\it superpotential} $W(x)$. The energy eigenvalues of the two Hamiltonians $H_\pm(x)$ are identical except for the ground state energy of $H_-(x)$: $E^{-}_{0} = 0,\; E^{+}_{n} = E^{-}_{n+1}$. The corresponding eigenfunctions are interrelated as,
 \begin{eqnarray}
 \psi^{+}_{n} (x) &=& (E^{-}_{n+1})^{-1/2} \,A\, \psi^{-}_{n+1} (x), \nonumber\\   
 \psi^{-}_{n+1} (x) &=& (E^{+}_n)^{-1/2} \, A^\dagger\, \psi^{+}_{n} (x),
 \end{eqnarray}
 where operators $A = \frac{d}{dx} + W(x)$ and $A^\dagger = -\frac{d}{dx} + W (x)$. The existence of a unique and normalizable ground state $\psi^{-}_0 (x)$ of $H_-(x)$ represents the unbroken phase of SUSY. The spectrum is bounded from below as $A\,\psi^{-}_0(x)=0$ implying that the superpotential determines the ground state: $\psi^{-}_0 (x) = exp(-\int^x W(x') dx')$. Though a particular $W(x)$ yields a definite pair of superpartners $V_\pm(x)$, for a given potential $V_-(x)$ an {\it isospectral deformation} of the type $W(x)\to{\tilde W} (x) = W(x) + g(x)$ is allowed subjected to the Bernoulli equation,
\begin{eqnarray}
g^2(x) + 2W(x) g(x) = g'(x).
\end{eqnarray}
However, the superpartner is now changed to ${\tilde V}_+(x)=V_+(x)+2g'(x)$ that still possesses a spectra identical to that of $V_+(x)$. On the other hand, this deformed superpotential does yield a different ground state for $V_-(x)$ and thus isospectral deformation effectively changes the boundary conditions for $V_-(x)$. It can be seen that boundaries of $\psi^{-}_0 (x)$ corresponds to the extrema of the superpotential.

\section{\label{sec:level3}Free particle in infinite box to KdV breathers}
The simplest quantum system of the free particle with $H = -d^2/dx^2$ possesses the eigenstates  $\psi_\kappa^{\pm}(x) = e^{\pm i \kappa x}$ and a spectrum: $E_\kappa = \kappa^2 > 0$. Imposing infinite boundaries (1-D box) at a distance $L$ apart fixes a particular superposition of $\psi_\kappa^{\pm}(x)$ for a given momentum $\kappa$ and discretizes the spectrum with a finite ground-state energy $a^2=\pi^2/L^2$. Since SUSY-QM requires $E_0^-=0$, on shifting to a non-zero potential $V_-(x)=-a^2$, Eq. \ref{E2} leads to the Riccati equation:
\begin{eqnarray}
W^2(x)-W'(x)=-a^2. 
\end{eqnarray}
The trivial solutions $W_\pm(x)=\pm\,ia$ supersymmetrically maps the potential to itself: $V_+(x)=-a^2=V_-(x)$ and the ground state would be a superposition of contributions $\exp(\mp iax)$ from both these superpotential. However, this clearly violates the imposed boundary conditions\footnote{Trivially, $W_\pm(x)=\pm\,ia$ represents the free particle without boundaries, as an energy scaling can always mean $a\to 0$.}. 

The non-trivial solutions to the Riccati equation are,
\begin{eqnarray}
\tilde{W}(x)=\begin{cases}
    a\tan\left(a\,x\right)\\
    -a\cot\left(a\,x\right)
\end{cases}.
\end{eqnarray}
Choosing the the boundaries at $x=0,\,L$ fixes $\tilde{W}(x)=-a\cot\left(a\,x\right)$. The eigenstates and eigenvalues turn out to be the standard ones. The superpotential embeds the knowledge of the boundaries and $V_-(x)$ has a distinct superpartner: $\tilde{V}_+(x)=-a^2+2a^2\csc^2\left(a\,x\right)$ \cite{khare,Khare2}. The isospectral deformation that connects the free particle and the 1-D box corresponds to the solution,
\begin{equation}
g(x) = - 2 ia \frac{\mathbf{e}^{2ia(x + c)}}{\mathbf{e}^{2ia(x + c)} - 1},\label{4}
\end{equation}
of Bernoulli’s equation with integration constant $c$. Retaining a non-vanishing $c$ amounts to,
\begin{eqnarray}
&&\tilde{W}(x)=-a\cot\,a(x+c),\nonumber\\
&&\tilde{V}_+(x)=-a^2+2a^2\csc^2a(x+c).
\end{eqnarray}

The boundary conditions of $V_-(x)$ ensures $c\in\mathbb{R}$. Noticeably, ${\tilde V}_+(x)$ has singularities at $x= l\,\pi-c$, representing a system of particles trapped within pairs of adjacent singularities where the wavefunctions vanish. We are motivated here to construct a general ${\tilde V}_+(x)$ by embedding $V_-(x)$ into the complex plane. More specifically, we are interested in complex potentials having applications in both quantum and optical systems where the $\mathcal{PT}$-symmetry plays a pertinent role in describing the various features like spectral singularities \cite{Longhi1}, unidirectional invisibility \cite{most1} and zero-width resonance \cite{ZA11} in the broken $\mathcal{PT}$-phase. Further, it is well-known that there is an intimate connection between isospectral potential and KdV as well as mKdV equations. Therefore, we also focus on establishing a similar relationship between the emergent potential and KdV equation in the present investigation. Keeping this in mind, we consider the integration constant $c$ to be complex in general\footnote{This is permissible as $c$ is not contained in the eigenvalues and in the worst case can be absorbed in the overall normalization.}. Then the boundary conditions $\psi_0^-(x=0)=0=\psi_0^-(x=L)$ can be generalized to,
\begin{eqnarray}
e^{-2ia\, x_{1,2}} = e^{2ia c},
\end{eqnarray}
with $x_{1,2}$ being the `new' {\it complex} locations of the boundaries. This further implies,
\begin{eqnarray}
x_2-x_1=\frac{m\pi}{a},\quad m\in\mathbb{Z}.
\end{eqnarray}
This is a 1-D infinite potential in the complex $x$ plane, placed parallel to the real axes, with \textit{real} `quantized' widths. The eigenvalues, however, are still real as they depend only $x_2-x_1$. One can consider the spectrum to belong to a definite width $\vert x_2-x_1\vert$ or as ground state energies of different wells of widths $\vert x_2-x_1\vert/m$. Thus, a \textit{family} of 1-D infinite wells are isospectral related to a family of complex potentials, which will be discussed next. 

\begin{figure}[ht]
\begin{center}
\includegraphics[scale=0.38]{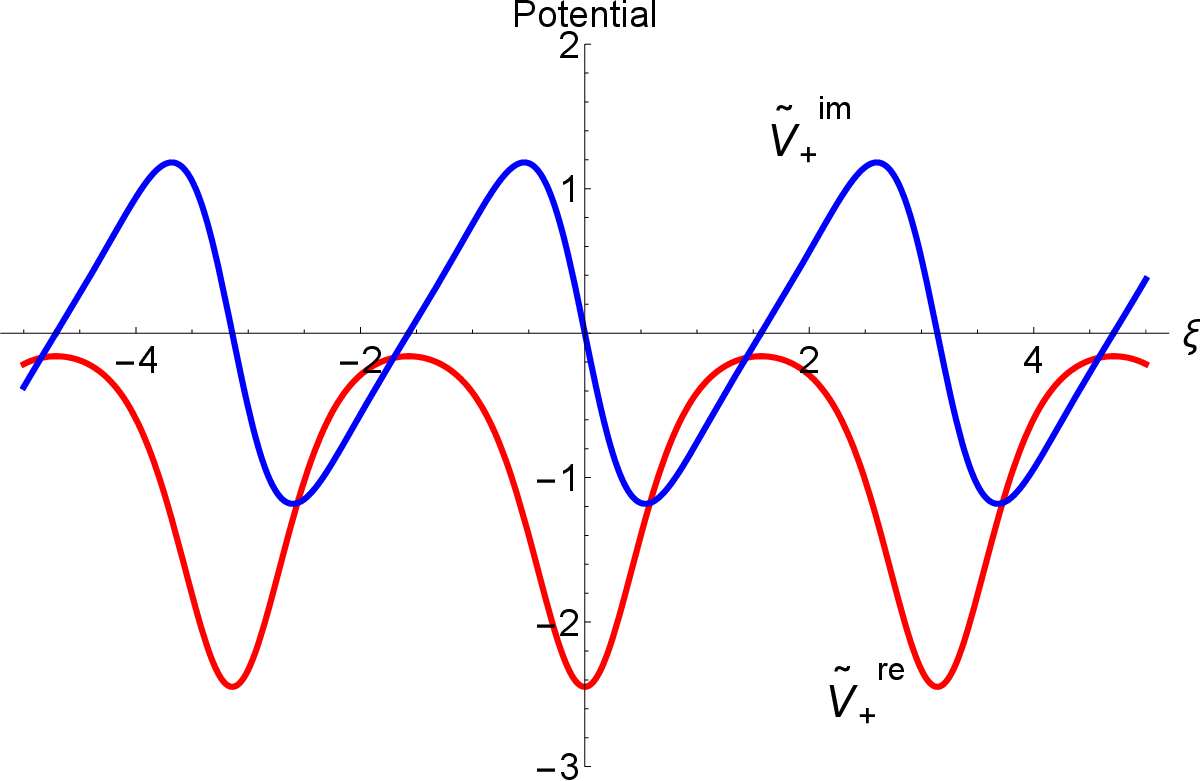}
\caption{Plots of  $\mathbb{R}e(\tilde{V}_+(x))$ and $\mathbb{I}m(\tilde{V}_+(x))$ vs. $\xi$.}
\label{fig1}
\end{center}
\end{figure}

\begin{figure}[ht]
\begin{center}
\includegraphics[scale=0.38]{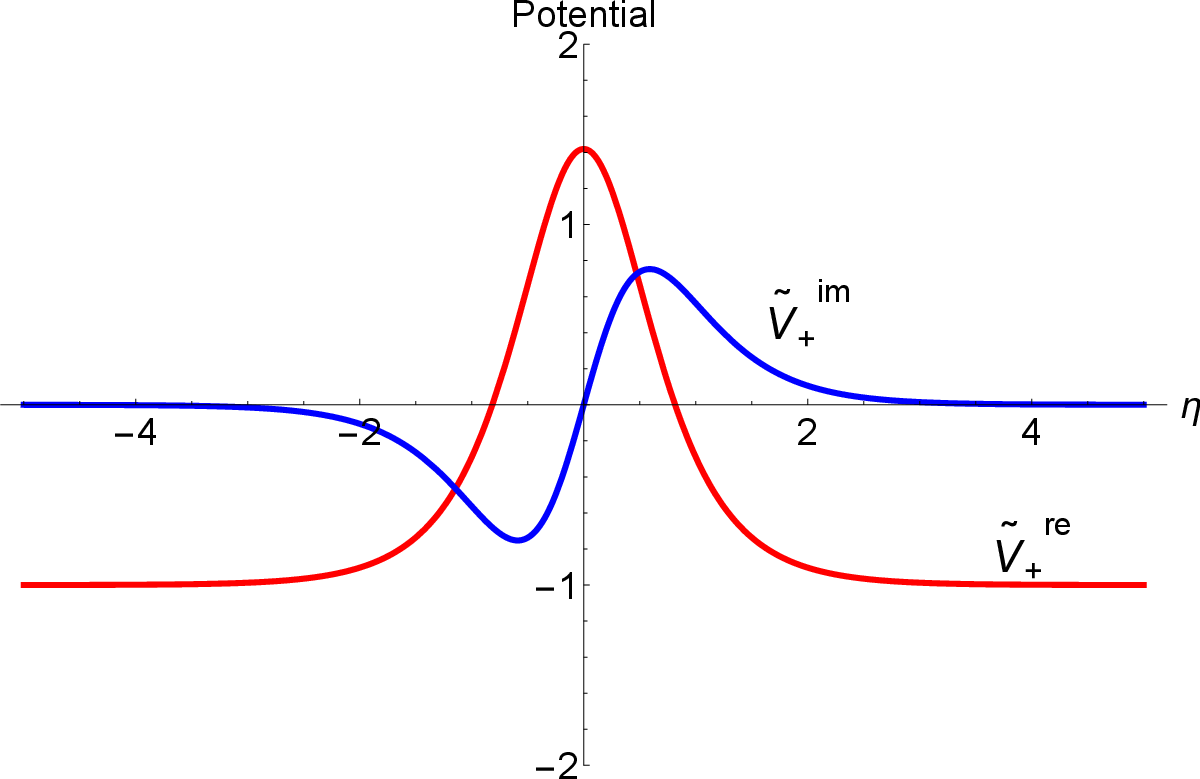}
\caption{$\mathbb{R}e(\tilde{V}_+(x))$ and $\mathbb{I}m(\tilde{V}_+(x))$ vs. $\eta$.}
\label{fig2}
\end{center}
\end{figure}

For mathematical and intuitive convenience we redefine the variable(s) as, 
\begin{eqnarray}
\xi &= a \left(x_{\rm re} + c_{\rm re}\right),\qquad
\eta &= a \left(x_{\rm im} +a c_{\rm im}\right),
\end{eqnarray}
with $x = x_{\rm re}+ i x_{\rm im}$. Then the isospectral partner potential can be represented as,
\begin{eqnarray}
{\tilde V}_+(x) = 
 a^2 \Bigg( \frac{4-4\,\cos \,2\xi \,\cosh\, 2\eta}{(\cos\, 2\xi - \cosh\, 2\eta)^2} -i \, 4\,\frac{\sin\, 2\xi\, \sinh \,2\eta}{(\cos \,2\xi - \cosh\, 2\eta)^2}-1\Bigg), \label{13}
\end{eqnarray}
where the real and imaginary parts of the potential are clearly separated in parameters $\xi$ and $\eta$. Fig. \ref{fig1} shows that ${\tilde V}_+^{\rm re}(x)$ is even and ${\tilde V}_+^{\rm im}(x)$ is odd $\xi$ which ensures $\mathcal{PT}$-symmetry of ${\tilde V}_+(x)$, provided we define parity transformation in the complex $(\xi,\eta)$-plane to be  $\xi\rightarrow -\xi,\, \eta\rightarrow\eta$ and time usual reversal as $i\rightarrow-i$\footnote{It is to be noted that the emergent $\mathcal{PT}$-symmetry is not manifest in terms of $x_{\rm re},\,x_{\rm im}$ due to the presence of $c$, the latter being necessarily non-zero for the `complexification' of the system.}. 
This further makes sense as the \textit{generalized} potential wells in the complex plane are manifested essentially through the shift of the origin. The variable  $\eta$ acts as a continuous parameter in $\{-\infty,\,\infty\}$, while $\xi$ serves as the coordinate restricted by boundary conditions for the well.
Actually in two dimensions, reflection of both the coordinates $(\xi\rightarrow -\xi,\, \eta\rightarrow -\eta)$ essentially amounts to a rotation by $\pi$-radians. A true (discrete) reflection amounts to either $(\xi\rightarrow -\xi,\, \eta\rightarrow\eta)$ or $(\xi\rightarrow \xi,\, \eta\rightarrow -\eta)$. We have opted for the first choice since the coordinate $\xi$ is real. Though the second choice also leaves $\tilde{V}_+(x)$ unchanged under ${\cal PT}$-transformation (Fig. \ref{fig2}), it cannot directly be related to the `physical motion' of the isospectral free particle motion. Also, since the ${\cal PT}$-invariant combination $i\eta$ will then also serve as the `coordinate', the results would be trivial.

\subsection{Complex KdV solutions}
In the context of KdV dynamics, the propagating solution can be described through a time parameter $t$, defined through $\tau = v\,t+x_0$ \cite{sukh}. For simplicity, we take $x_0 =0$, which  retains the symmetric nature of potential at $t=0$. As mentioned in \cite{sukh}, setting $\xi = aX+ 4a^3\,t$ in Eq. (\ref{13}) with arbitrary $\eta$, the isospectral potential ${\tilde V}_+(x)$ is an exact solution of the defocusing-type KdV equation,
\begin{eqnarray}
u_t - 6\,u\, u_X + u_{XXX} =0. \label{KdV}
\end{eqnarray}
Under these parametric identifications, potential ${\tilde V}_+(x)$ is periodic in both space and time representing the characteristics of breather \cite{zaw}. It is worth mentioning that, breathers have been observed experimentally in Bose-Einstein condensates \cite{smerzi} and optical waveguides \cite{kutz}.
Recently, a similar kind of solution has also been shown to exist for focusing-type KdV equation $u_t+6uu_X+u_{XXX}=0$ with complex soliton solution \cite{cen}, where the time-delay analysis has been done \cite{correa}. ${\tilde V}_+(x)$ also reproduces this complex soliton under the identification $\eta=-aX+4a^3t$ while $\xi$ being arbitrary now. Remarkably, this behavior is synonymous with the motion of the free particle in the complex plane decaying along the imaginary direction away from the box.

\begin{figure}[bt]
    \centering
    \begin{minipage}{0.24\textwidth}
        \centering
        \includegraphics[width=\linewidth]{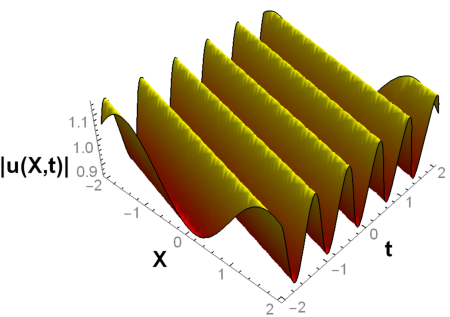}
        (a)
    \end{minipage}%
    \begin{minipage}{0.24\textwidth}
        \centering
        \includegraphics[width=\linewidth]{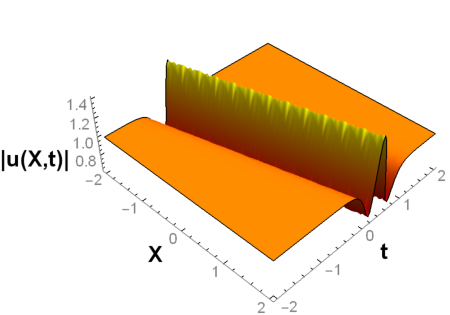}        
        (b)
    \end{minipage}
    \caption{KdV solutions of class I: (a) Complex breather for $\xi=aX+4a^3t$ with $\eta=2$ and (b) complex soliton for $\eta=-aX+4a^3t$ with $\xi=2$. In both cases $a=1$.}
    \label{Fig3}
\end{figure}

Additionally, $\tilde{V}_+$ further satisfies the defocussing KdV equation for the parameterization $\xi=aX-2a^3t$ with $\eta$ arbitrary, and its focusing counterpart for $\eta=aX+2a^3t$ with $\xi$ arbitrary. These two cases correspond to breather and complex soliton behaviors, respectively. The corresponding superpotential $\tilde{W} = -a \,\cot(\xi+i\eta)$ also then satisfies the defocusing and focusing-type mKdV equations \cite{buti, miura2}
\begin{eqnarray}
v_{1,t} - 6v^2_1 v_{1,X}+v_{1,XXX} =0, \nonumber\\
v_{2,t} + 6v^2_1 v_{2,X}+v_{2,XXX} =0,
\end{eqnarray}
respectively. Indeed, the Riccati equations relating the potential and superpotential are identified as the corresponding Miura transformations \cite{miura1}:
\begin{eqnarray}
u=v_1^2\pm v_{1,X}=v_2^2\pm iv_{2,X},
\end{eqnarray}
since the two types of mKdV solutions are related as $v_2=iv_1$ \cite{buti, MSP}.
Interestingly, for $4a^3$ being the coefficient of $t$, the Miura transformation equation does not possess any solution and thus $\tilde{W}$ does {\it not} satisfy any mKdV equation. Therefore we have obtained two distinct classes of KdV solutions from the isospectral potential. We summarize these results in table \ref{Table1} and figures \ref{Fig3} and \ref{Fig4} depict some examples of these two classes of periodic and localized complex solutions.

In the next section, we generalize the superpotential through parametric extensions leading to both broken and unbroken phases of $\mathcal{PT}$. The $\mathcal{PT}$-broken phase spectrum further supports a subset that are zero-width resonances having real energy, given certain algebraic criteria are met.

\begin{center}
\begin{table}[bt]
\caption{Conditions for obtaining KdV and mKdV solutions}
\label{Table1}
\begin{NiceTabular}{|c||c|c|c|c| }
\hline
{\rm Class}&$\xi$&
$\eta$&
\makecell{$\tilde{V}_+$}&
\makecell{$\tilde{W}$}\\
\hline\hline
\multirow{2}{*}{\rm I}&$aX+4a^3t$ & {\rm Parameter} & \makecell{Breather for Defocussing\\KdV} & {\rm None}\\
\cline{2-5}
&{\rm Parameter} & $-aX+4a^3t$ & \makecell{Soliton for Focusing\\KdV} & {\rm None}\\
\hline
\multirow{2}{*}{\rm II}&$aX-2a^3t$ & {\rm Parameter} & \makecell{Breather for Defocussing\\KdV} & \makecell{Breather for Defocussing\\mKdV}\\
\cline{2-5}
&{\rm Parameter} & $aX+2a^3t$ & \makecell{Soliton for Focusing\\KdV} & \makecell{Soliton for Focusing\\mKdV}\\
\hline
\end{NiceTabular}
\end{table}
\end{center}

\section{\label{sec:level5} Generalized complex potential: ${\cal PT}$-unbroken and broken phases}
For a general complex potential $V(\xi, \eta)$ on the $\xi$-$\eta$ plane, the Schr\"odinger eigenvalue equation is given as,
\begin{eqnarray}
\Big(\frac{\partial^2}{\partial \xi^2} + \frac{\partial^2}{\partial \eta^2} -   V(\xi,\eta) \Big)\, \Psi(\xi,\eta) = - k^2 \, \Psi(\xi,\eta),
\end{eqnarray}
with energy $E=k^2$. Considering a particular solution of the form: $\Psi(\xi,\eta) =  \Phi(\xi,\eta) \, e^{i{\tilde k}\eta}$, with a complex momentum ${\tilde k}$, results in,
 \begin{eqnarray}
\frac{1}{\Phi(\xi,\eta)}\Big(\frac{\partial^2}{\partial \xi^2} -  V(\xi,\eta) \Big)\Phi(\xi,\eta) + k^2 = {\tilde k}^2. \label{17}
\end{eqnarray}
Taking $V=0$ in the case of the infinite well and imposing the boundary conditions, $\Psi (\xi_1, \eta_1) = 0=\Psi (\xi_2, \eta_2)$, the particle is confined to move along the real axis $\xi$ yielding a real spectrum. In the orthogonal direction, the physical variable is $i\eta$ instead of $\eta$ which corresponds to an imaginary momentum affecting an exponential decay of the eigenstates. Consequently, as the complex $\mathcal {PT}$-symmetric potential ${\tilde V}_+(\xi,\eta)$ is isospectral to this real system, the prior must be confined to its $\mathcal {PT}$-symmetric phase. The corresponding eigenfunctions further ensure this, 
 
\begin{eqnarray}
&&\psi^+_n (\xi,\eta) \propto (n+1)\,\Bigg[\cos\{(n+2)\xi\}\cosh\{(n+2)\eta\}-i\sin\{(n+2)\xi\}\sinh\{(n+2)\eta\}\Bigg]\nonumber\\
&&\qquad\qquad\quad-\frac{\cos\{(n+1)\xi\}\sinh\{(n+1)\eta\}- i\sin\{(n+1)\xi\}\cosh\{(n+1)\eta\} }{\big(\sin\xi\,\cosh \eta 
- i \,\cos \xi\, \sinh \eta \big)},
\end{eqnarray}
which are $\mathcal{PT}$-symmetric in $\xi$. Additionally, a single superpotential ${\tilde W}(\xi,\eta)$ generates this whole spectrum, which is a defining  characteristic of an isospectral potential with unbroken $\cal{PT}$-symmetry \cite{9} unless any additional symmetry is involved \cite{BQ01}, having a vanishing `current' \cite{KAP}. Along the real direction, with appropriate boundary restrictions, these eigenfunctions are sinusoidal whereas they exponentially decay along the imaginary direction. If alternate boundary conditions can change this decaying nature, it should lead to a non-zero associated current \cite{KAP} and thereby to the $\mathcal{PT}$-broken phase.

Just like ${\tilde V}_+(\xi,\eta)$, eigenstates $\psi_n^+(\xi,\eta)$ are also $\mathcal{PT}$-symmetric under the alternate parity choice: $\xi\rightarrow\xi,\,\eta\rightarrow-\eta$ that we had avoided on physical grounds. Evidently, within no range of the present parameters, the superpartner can display spontaneous breaking of $\cal{PT}$-symmetry. Naturally, there is a more general form of ${\tilde V}_\pm(\xi,\eta)$ which reduces to that in Eq. (\ref{13}) under certain parametric conditions preserving $\cal{PT}$-symmetry \cite{9}. A simple extension could be when $\tilde{W}(\xi,\eta)$ can be generalized to {\it two} superpotentials $\tilde{W}_1(\xi,\eta)=-\left(a\pm i\gamma\right)\cot\left(\xi+i\eta\right)$, with $a,\alpha,\gamma\in\mathbb{R}$ and further $\xi=\alpha x_{\rm re}+\alpha c_{\rm re}$ and $\eta=\alpha x_{\rm im}+\alpha c_{\rm im}$. This equivalently amounts to a parametrically more generalized ground state. The corresponding generalized supersymmetric potentials,
\begin{eqnarray}
\tilde{V}^1_-(x)=(a\pm i\gamma)(a\pm i\gamma-\alpha)\csc^2\left(\xi+i\eta\right),
\end{eqnarray}
up to constant energy, are connected to the free particle under certain parametric conditions \cite{mat1,mat2,mat3}. These potentials are shape-invariant \cite{ged, khare} under the transformation $a\rightarrow a+\alpha$. There are indeed two superpotentials, which are mandatory for $\mathcal{PT}$-symmetry breaking \cite{9}, and the two potentials go to a {\it unique} one on imposition of $\mathcal{PT}$-symmetry condition: $\gamma(2a-\alpha)=0$. The $\mathcal{PT}$-symmetric phase corresponds to $\gamma=0$ with real energies $E^{1,{\rm s}}_n=(a+n\alpha)^2$ obtained through shape invariance. In particular, $V^1_-(x)$ represents a free particle for $\alpha=a$. The broken phase has $a=\alpha/2$ with a complex-conjugate paired spectrum,
\begin{eqnarray}
E^{1,{\rm b}}_n=\left(n\alpha+\frac{\alpha}{2}\pm i\gamma\right)^2.
\end{eqnarray}
This spectrum is always complex and does not allow for exceptional points since $n=-1/2$ is not possible. It is worth noting that the isospectrality to any real system is not possible in this sector. Further confirmation comes from the eigenstates of the system,
\begin{eqnarray}
\psi_n^{1,{\rm b}}(y)&\propto&\left(-\frac{d}{dy}+\tilde{W}_1(y)\right)^n\left(\sin(y)\right)^{a+n\alpha\pm i\gamma},\nonumber\\
y&=&\xi+i\eta,\label{NEq001}
\end{eqnarray}
that, although lead to a non-vanishing current \cite{KAP}, cannot result in a vanishing Wronskian unless $\gamma=0$ (symmetric phase). 

In order to obtain a more general $\mathcal{PT}$-symmetric system having zero-width resonances, we further generalize to,
\begin{eqnarray}
\tilde{W}_2(\xi,\eta)=-a\cot\left(\xi+i\eta\right)+(b\pm i\lambda)\tan\left(\xi+i\eta\right),\quad a,b,\lambda&\in&\mathbb{R}.
\end{eqnarray}
The $\mathcal{PT}$-symmetry is imposed through the condition $\lambda(2b-\alpha)=0$. The symmetric phase with $\lambda=0$ corresponds to a unique superpotential that leads to the potential and the subsequent spectrum:
\begin{eqnarray}
&&\tilde{V}^{2,{\rm s}}_-(\xi,\eta)=a(a-1)\csc^2\left(\xi+i\eta\right)+b(b-\alpha)\sec^2\left(\xi+i\eta\right),\nonumber\\
&&E^{2,{\rm s}}_n=(a+b+2n\alpha)^2,
\end{eqnarray}
the latter obtained through shape-invariance shifts $a\to a+\alpha,\quad b\to b+\alpha$. This potential is the trigonometric analog to the P\"oschl-Teller II potential. The broken phase requires $b=\alpha/2$, which maintains two superpotentials, still leading to a unique potential and subsequent spectrum:
\begin{figure}[bt]
    \centering
    \begin{minipage}{0.24\textwidth}
        \centering
        \includegraphics[width=\linewidth]{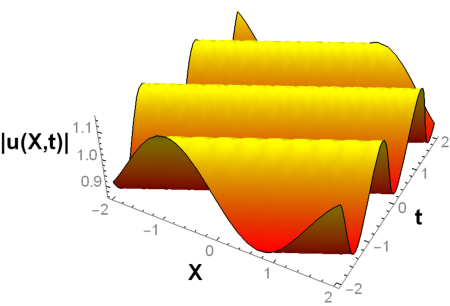}
        (a)
    \end{minipage}%
    \begin{minipage}{0.24\textwidth}
        \centering
        \includegraphics[width=\linewidth]{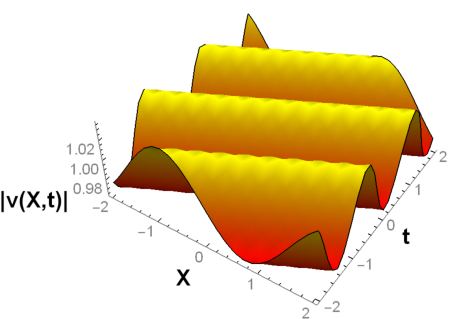}        
        (b)
    \end{minipage}
    \begin{minipage}{0.24\textwidth}
        \centering
        \includegraphics[width=\linewidth]{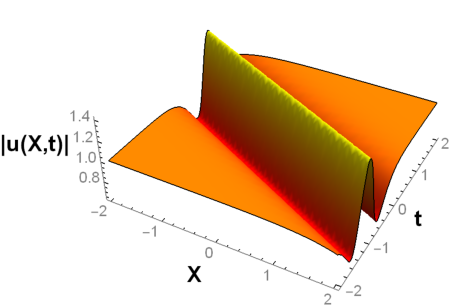}
        (c)
    \end{minipage}%
    \begin{minipage}{0.24\textwidth}
        \centering
        \includegraphics[width=\linewidth]{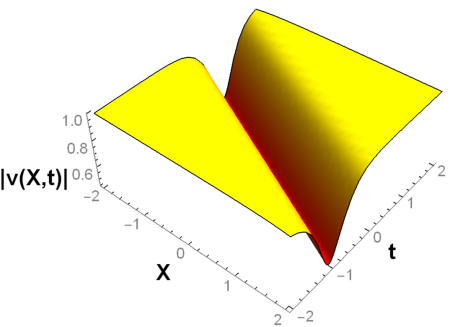}        
        (d)
    \end{minipage}
    \caption{Solutions of class II: (a) Complex KdV and (b) complex mKdV breathers for $\xi=aX-2a^3t$ with $\eta=2$. (c) Complex KdV and (d) complex mKdV solitons for $\eta=aX+2a^3t$ with $\xi=2$. In all cases $a=1$.}
    \label{Fig4}
\end{figure}
\begin{eqnarray}
&&\tilde{V}^{2,{\rm b}}_-(\xi,\eta)=a(a-1)\csc^2\left(\xi+i\eta\right)-\left(\lambda^2+\frac{1}{4}\right)\sec^2\left(\xi+i\eta\right),\nonumber\\
&&E^{2,{\rm b}}_n=\left(a+\frac{\alpha}{2}+2n\alpha\pm i\gamma\right)^2.
\end{eqnarray}
Clearly, zero-width resonances are now possible for parametric exceptional points $a=-2n\alpha-\alpha/2$, consistent with the fact that $a$ must represent free-particle ground state momentum in the symmetric phase. The eigenstates in the broken sector,
\begin{eqnarray}
\psi_n^{2,{\rm b}}(y) \propto \left(-\frac{d}{dy}+\tilde{W}_2(x)\right)^n\left(\sin(y)\right)^{a+n\alpha}\left(\cos(y)\right)^{\frac{\alpha}{2}+n\alpha\pm i\lambda},
\end{eqnarray}
also correspond to non-vanishing current, and more interestingly, to vanishing Wronskians for $a=-2n\alpha-\alpha/2$. Here, it is worth noting that the choice of the ground state, or equivalently the choice of the superpotential, plays a key role in capturing the $\mathcal{PT}$-broken phase. Further, the connection to the KdV equation is no longer possible given the isospectrality with the free particle no longer exists.

To obtain non-trivial zero-width resonances the $\tan(\xi+i\eta)$ term in $\tilde{W}_2(\xi,\eta)$ is necessary. Its importance follows from the $sl\left(2,\mathbb{R}\right)$ algebraic structure \cite{Gangopadhyaya01} obeyed by the isospectral potentials\footnote{For real potentials with sinusoidal functions it is $SO(3)$. For potentials with hyperbolic functions, it is $SO(2,1)$ for real ones and $sl\left(2,\mathbb{C}\right)$ for $\mathcal{PT}$-symmetric ones \cite{BQ01}.}. Though $\tilde{V}^{2,{\rm b}}_-(\xi,\eta)$ does not satisfy this algebra directly, it can be re-arranged as a trigonometric analog of the P\"oschl-Teller potential that does \cite{BQ01}:
\begin{eqnarray}
&&\frac{1}{4}\left(\Lambda-\Gamma\right)\left(\Lambda-\Gamma-\alpha\right)\csc^2(y)+\frac{1}{4}\left(\Lambda+\Gamma\right)\left(\Lambda+\Gamma+\alpha\right)\sec^2(y)\nonumber\\
&&\qquad\qquad=\left[\Lambda^2+\Gamma\left(\Gamma+\alpha\right)\right]\csc^2(2y)-\Lambda\left(2\Gamma+\alpha\right)\csc(2y)\cot(2y),\quad\Gamma,\Lambda\in\mathbb{R}.\label{EQa}
\end{eqnarray}
wherein the coefficients should be consistent with those of $\tilde{V}^{2,{\rm b}}_-(y)$. The potential form on the RHS of Eq. \ref{EQa} satisfies the $sl\left(2,\mathbb{R}\right)$ potential algebra and corresponds to a superpotential $\bar{W}(y)=\Gamma\cot(2y)-\Lambda\csc(2y)$. In terms of the parameters of the algebra, the potential takes the form,
\begin{eqnarray}
V_m(z)=\left(\frac{1}{4}-m^2\right)\frac{dF(z)}{dz}+2m\frac{dG(z)}{dz}+G(z)^2,\quad
z=2y,
\end{eqnarray}
that enables the identification,
\begin{eqnarray}
\bar{W}(z)=\begin{cases}
    {\rm either}\quad\left(\mp m-\frac{1}{2}\right)F(z)\pm G(z)\\
    {\rm or}\qquad\quad\left(\pm\beta-\frac{1}{2}\right)F(z)\pm\frac{m}{\beta}G(z).
\end{cases}
\end{eqnarray}
The two functions $F(z)=\cot(z)$ and $G(z)=\beta\csc(z)$ are a particular $sl\left(2,\mathbb{R}\right)$ representation which need satisfy,
\begin{eqnarray}
\frac{dF}{dz}=-1-F^2\quad{\rm and}\quad \frac{dG}{dz}=-FG.\label{NEq1}
\end{eqnarray}
The complete algebra is represented as\footnote{The $sl\left(2,\mathbb{C}\right)$ counterpart of this was obtained in references \cite{Gangopadhyaya01,BQ01}.},
\begin{eqnarray}
&&\left[J_0\,,\,J_\pm\right]=\pm J_\pm,\quad\left[J_+\,,\,J_-\right]=2J_0;\nonumber\\
&&J_0=\frac{\partial}{\partial_\varphi},\nonumber\\
&&J_\pm=e^{\pm\varphi}\left[\pm\frac{\partial}{\partial z}+\left(-\frac{\partial}{\partial\varphi}\mp\frac{1}{2}\right)F(z)+G(z)\right],
\end{eqnarray}
with the Casimir $J^2=J_0^2-J_0+J_+J_-$. The corresponding spectrum is,
\begin{eqnarray}
&&J^2\vert k,m\rangle=k(k-1)\vert k,m\rangle,\quad J_0\vert k,m\rangle=m\vert k,m\rangle,\nonumber\\
&&m=-k,-k+1,\cdots,k.
\end{eqnarray}

It is clear that $\tilde{V}^{1,{\rm b}}_-(\xi,\eta)$ corresponds to a reduced $sl\left(2,\mathbb{R}\right)$ representation with $F(y)=\cot(y)$ and $G(y)=0$. Indeed the underlying potential algebra for a potential of $\csc^2(\xi+i\eta)$-type is $SU(1,1)$ \cite{Alhassid01}, which is isomorphic to $sl\left(2,\mathbb{R}\right)$. For $\mathcal{PT}$-symmetry all the coefficients in Eq. \ref{EQa} must be real as $y\to-y$ under $\mathcal{PT}$, which is consistent with the fact that $\Gamma,\Lambda\in\mathbb{R}$. This considerably restricts the parameters $a(\Gamma,\Lambda)$, $b(\Gamma,\lambda)$ and $\lambda(\Gamma,\lambda)$ and it follows that $b,\lambda\neq0$ for $a\in\mathbb{R}$. In other words, since $a$ is the real momentum of the 1-D infinite well, a sufficiently general $\mathcal{PT}$-symmetric structure that leads to non-trivial zero-width resonances demands that $\tilde{W}_2(x)$ carries the $\tan(\xi+i\eta)$ term. A simpler alternative could have been to extend $\tilde{W}(x)$ to:
\begin{eqnarray}
\tilde{W}_3(\xi,\eta)=-a\cot\left(\xi+i\eta\right)+{\cal B}\csc\left(\xi+i\eta\right),\quad a\in\mathbb{R},
\end{eqnarray}
which directly provides a set of non-trivial $F$ and $G$ satisfying $sl\left(2,\mathbb{R}\right)$. However, the existence of a $\mathcal{PT}$-broken phase requires more than ${\cal B}\in \mathbb{C}$ since the corresponding spectrum $E^{3,b}_n=(a+n\alpha)^2$ does not feature ${\cal B}$; which would correspond to a more `drastic' change in parameterization than that for $\tilde{W}_2(x)$ {\it e. g.} $a\in \mathbb{C}$. The potential algebra for this system (P\"oschl-Teller II potential) is actually $SO(2,2)$ \cite{Wu01} which is not directly connected to $sl(2,\mathbb{R})$. This further strengthens the notion that non-trivial $sl\left(2,\mathbb{R}\right)$ (or $sl\left(2,\mathbb{C}\right)$ in the hyperbolic case) representation is important to obtain zero-width resonances in the $\mathcal{PT}$-broken phase.

\section{\label{sec:level6}Conclusion}

In conclusion, the isospectral deformation of the free particle enables a natural generalization of the quantum system to complex potentials, through a more general implementation of the boundary conditions that extends to the complex plane. The particle in a box case leads to the complex P\"oschl-Teller type potential, defined on the complex plane, where the physical coordinate is associated with a non-compact parameter. The underlying geometry of space has an interpretation, with periodic wavefunctions along the real coordinate, that is endowed with parity and time-reversal symmetry. The case of generalized complex potential also arises along this real coordinate, in the underlying complex plane, with distinctly different boundary behavior. In the process, the complex potential is found to satisfy the KdV equation, with the real and imaginary components corresponding to breather and localized dynamics respectively, possessing an emergent $\mathcal{PT}$-symmetry. The breather and complex soliton appear with appropriate parametric interpretations. Furthermore, the generalization of the superpotential leads to the shape-invariant partner potentials, which provides a natural connection of two forms of Miura transforms $u = v^2 \pm v_x$ and $u=-v^2 \pm iv_x$, further satisfying the mKdV equation. In addition, a proper and generalized choice of the ground state having the complex energy brings out the $\mathcal{PT}$-broken sector of the system given the $sl(2,\mathbb{R})$ potential algebra is satisfied, where the states with zero-width resonance are found to occur. This further provides a physical realization of the resonances in the broken-$\mathcal{PT}$ phase from a manifest algebraic basis.

In the future, the KdV hierarchy of the isospectral potential connected to the free particle within the framework of $\mathcal{PT}$-symmetry is worth exploring, with their connection to multi-solitons \cite{sukh}. Since the complex KdV system is also related to the Levi-Civita equation \cite{levi1, NN01}, it would be interesting to study the latter in the view of ${\cal PT}$-symmetric systems. In addition, the connection of $\mathcal{PT}$-breaking and the corresponding potential algebra can be investigated further. Moreover, the Berry phase correlated with the complex solution in the $\mathcal{PT}$-broken  phase, is under investigation and will be reported elsewhere.

\section*{Acknowledgment}
Kumar Abhinav’s research is supported by Mahidol University, Thailand under the Grant Number MRC-MGR 04/2565. Aradhya Shukla and Prasanta K Panigrahi acknowledge the support from DST, India through Grant No.: DST/ICPS/QuST/Theme-1/2019/2020-21/01.

\let\doi\relax

\end{document}